\documentstyle[preprint,pra,aps,fleqn]{revtex}

\begin{document}
\draft
\title{Two-mode Nonlinear Coherent States}
\author{
Xiao-Guang Wang\thanks{%
email:xyw@aphy.iphy.ac.cn, or xiao-guang.wang@physik.tu-chemnitz.de,
Recieved 31. 1. 2000,
Revised 30. 3. 2000,
Accepted 2. 4. 2000 in Optics Communications by Wolfgang P. Schleich,
Editor, Editor's number: WS 1744}
}
\address{Chinese Center of Advanced Science and Technology (World Laboratory),\\
P.O.Box 8730, Beijing 100080 \\
and Laboratory of Optical Physics,  Institute of Physics, Chinese Academy of%
\\
Sciences,Beijing, 100080, P. R. China}
\date{\today}
\maketitle

\begin{abstract}
Two-mode nonlinear coherent states are introduced in this paper. The pair
coherent states and the two-mode Perelomov coherent states are special cases
of the two-mode nonlinear coherent states. The exponential form of the
two-mode nonlinear coherent states is given. The photon-added or
photon-subtracted two-mode nonlinear coherent states are found to be 
two-mode nonlinear coherent states with different nonlinear functions. The
parity coherent states are introduced as examples of two-mode nonlinear
coherent states, and they are superpositions of two corresponding coherent
states. We also discuss how to generate the parity coherent states in the
Kerr medium.
\end{abstract}


\section{Introduction}

Coherent states of a simple harmonic oscillator\cite{CS} have considerable
applications in the study of quantum optics. Recently another type of
coherent states, nonlinear coherent states (NLCSs)\cite{NLCS1,NLCS2} as well
as their superpositions\cite{SUPERNLCS}, have been introduced and studied.
The so-called nonlinear coherent states are defined as the right-hand
eigenstates of the product of the boson annihilation operator $a$ and a
non-constant function of number operator $\hat{N}=a^{\dagger }a,$ 
\begin{equation}
f(\hat{N})a|\alpha ,f\rangle =\alpha |\alpha ,f\rangle ,
\end{equation}
where $f(\hat{N})$ is an operator-valued function of the number operator and 
$\alpha $ is a complex eigenvalue. The ordinary coherent states $|\alpha
\rangle $ are recovered for the special choice of $f(\hat{N})=1.$ A class of
NLCSs can be realized physically as the stationary states of the
center-of-mass motion of a trapped ion\cite{NLCS1}. These NLCSs exhibit
various non-classical features like squeezing and self-splitting. The notion
of the NLCS was generalized to the two-photon NLCS\cite{TPNLCS}. One
two-photon NLCS is the squeezed vacuum state and another is squeezed first
Fock state\cite{TPNLCS,wang2}. We want to generalize the notion of the NLCS
to two-mode case in this paper.

In a previous paper\cite{wang1}, we have given an exponential form of
one-mode NLCS $|\alpha ,f\rangle $ and proved that photon-added one-mode
NLCSs are still NLCSs. In analogy to the defination of the one-mode NLCS,
the two-mode nonlinear coherent state(TMNLCS) is defined as

\begin{equation}
f(\hat{N}_a,\hat{N}_b)ab|\alpha ,f,q\rangle =\alpha |\alpha ,f,q\rangle ,
\end{equation}
where $a$ and $b$ are boson annihilation operators; $f(\hat{N}_a,\hat{N}_b)$
is the function of the number operator $\hat{N}_a=a^{\dagger }a$ and $\hat{N}%
_b=b^{\dagger }b;$ $q$ is the photon number difference between two modes of
the field. In section II, we give the expansion and exponential form of the
TMNLCS. In section III and IV, we discuss the photon-added TMNLCS and
photon-subtracted TMNLCS. Section V introduces the parity pair coherent
states and parity two-mode Perelomov cohernet states as interesting examples
of the TMNLCSs. A conclusion is given in section VI.

\section{Expansion and exponential form of the TMNLCS.}

We next determine the solution to the eigenvalue equation (Eq.(2)). The
TMNLCS has the form

\begin{equation}
|\alpha ,f,q\rangle =\sum_{n=0}^\infty C_n|n+q,n\rangle .
\end{equation}
Substituting Eq.(3) into Eq.(2), we find the recursion relation among the
coefficients $C_n$'s

\begin{eqnarray}
C_{n+1} &=&\frac \alpha {f(n+q,n)\sqrt{(n+1)(n+q+1)}}C_n, \\
C_n &=&\alpha ^n\sqrt{\frac{q!}{n!(n+q)!}}\left[ \prod_{m=1}^n\frac 1{%
f(m+q-1,m-1)}\right] C_0.
\end{eqnarray}
Thus the expansion of the TMNLCS is obtained as 
\begin{eqnarray}
|\alpha ,f,q\rangle  &=&C_0\sum_{n=0}^\infty \alpha ^n\sqrt{\frac{q!}{%
n!(n+q)!}}\left[ \prod_{m=1}^n\frac 1{f(m+q-1,m-1)}\right] |n+q,n\rangle  \\
&=&C_0\sum_{n=0}^\infty \frac{\alpha ^nq!a^{\dagger n}b^{\dagger n}}{n!(n+q)!%
}\left[ \prod_{m=1}^n\frac 1{f(m+q-1,m-1)}\right] |q,0\rangle .  \nonumber
\end{eqnarray}
One can show that 
\begin{equation}
\lbrack g(\hat{N}_a,\hat{N}_b)a^{\dagger }b^{\dagger }]^n=a^{\dagger
n}b^{\dagger }{}^n\prod_{m=1}^ng(\hat{N}_a+m,\hat{N}_b+m).
\end{equation}
Here $g(\hat{N}_a,\hat{N}_b)$ is an arbitrary function of $\hat{N}_a$ and $%
\hat{N}_b.$ Then using Eq.(7) with 
\begin{equation}
g(\hat{N}_a,\hat{N}_b)=\frac \alpha {f(\hat{N}_a-1,\hat{N}_b-1)\hat{N}_a},
\end{equation}
the state $|\alpha ,f,q\rangle $ is finally written in the exponential form 
\begin{eqnarray}
|\alpha ,f,q\rangle  &=&C_0\sum_{n=0}^\infty \frac{[g(\hat{N}_a,\hat{N}%
_b)a^{\dagger }b^{\dagger }]^n}{n!}|q,0\rangle   \nonumber \\
&=&C_0\exp [g(\hat{N})a^{\dagger }b^{\dagger }]|q,0\rangle   \nonumber \\
&=&C_0\exp [\frac \alpha {f(\hat{N}_a-1,\hat{N}_b-1)\hat{N}_a}a^{\dagger
}b^{\dagger }]|q,0\rangle .
\end{eqnarray}
Here, $C_0$ can be determined as 
\begin{equation}
C_0=\left\{ \sum_{n=0}^\infty \frac{q!|\alpha |^{2n}}{n!(n+q)!}\left[
\prod_{m=1}^n\frac 1{f(m+q-1,m-1)}\right] ^2\right\} ^{-1/2}.
\end{equation}

In fact, by direct verification, we have 
\begin{equation}
\lbrack f(\hat{N}_a,\hat{N}_b)ab,\frac 1{f(\hat{N}_a-1,\hat{N}_b-1)\hat{N}_a}%
a^{\dagger }b^{\dagger }]=1.
\end{equation}
From the above commutaion relations and the definition of the TMNLCS, we can
also obtain the exponential form of the TMNLCS. Next we want to give some
examples of the TMNLCS.

The pair coherent state is an important correlated two-mode field. It is
defined as\cite{pair}

\begin{equation}
ab|\zeta ,q\rangle =\zeta |\zeta ,q\rangle ,
\end{equation}
which is the state $|\alpha ,f,q\rangle $ with  $f(\hat{N}_a,\hat{N}%
_b)\equiv 1.$ Then from Eq.(2), (9) and (12), the exponential form of the
pair coherent state is

\begin{equation}
|\zeta ,q\rangle =\exp (\frac \zeta {\hat{N}_a}a^{\dagger }b^{\dagger
})|q,0\rangle 
\end{equation}
up to a normalization constant.

Another important correlated two-mode field is the two-mode Perelomov
coherent state, which is defined as\cite{Perelomov}

\begin{equation}
|\xi ,q\rangle =\exp (\xi a^{\dagger }b^{\dagger }-\xi ^{*}ab)|q,0\rangle
\end{equation}
We find that the state $|\xi ,q\rangle $ satisfies the equation\cite{wang2}

\begin{equation}
\frac 2{\hat{N}_a+\hat{N}_b+q+2}ab|\xi ,q\rangle =\frac{\xi \tanh (|\xi |)}{%
|\xi |}|\xi ,q\rangle ,
\end{equation}
which shows that two-mode Perelomov coherent states are TMNLCSs with the
nonlinear function $2/(\hat{N}_a+\hat{N}_b+q+2).$ In the subspace $F_q\equiv 
$span$\{|n+q,n\rangle |n=0,1,2...\}$ of the two-mode Fock space, Eq.(15) can
be simplified as 
\begin{equation}
\frac 1{\hat{N}_a+1}ab|\xi ,q\rangle =\frac{\xi \tanh (|\xi |)}{|\xi |}|\xi
,q\rangle ,
\end{equation}

Then From Eq.(9) , the exponential form of the state $|\xi ,q\rangle $ is
obtained as

\begin{equation}
|\xi ,q\rangle =\exp [\frac{\xi \tanh (|\xi |)}{|\xi |}a^{\dagger
}b^{\dagger }]|q,0\rangle 
\end{equation}
up to a normalization constant. Actually using the well-known identity

\begin{eqnarray}
&&\exp (\xi a^{\dagger }b^{\dagger }-\xi ^{*}ab)  \nonumber \\
&=&\exp [\frac{\xi \tanh (|\xi |)}{|\xi |}a^{\dagger }b^{\dagger }][\frac 1{%
\cosh (|\xi |)}]^{\hat{N}_a+\hat{N}_b+1}\exp [\frac{-\xi ^{*}\tanh (|\xi |)}{%
|\xi |}ab],
\end{eqnarray}
one can directly verifiy Eq.(17).

One special type of two-mode Perelomov coherent state, $|\xi ,0\rangle ,$ is
the eigenstate of the two-mode phase operator $1/\sqrt{(1+\hat{N}_a)(1+\hat{N%
}_b)}ab$ (also called two-mode squeezed vacuum state)\cite{Tmphase}. The
eigenstate of the two-mode phase operator seems different from the state $%
|\xi ,0\rangle .$ However, since $q=0,$ one can show that the operator $1/(1+%
\hat{N}_a)ab$ is identical to the two-mode phase operator in the subspace $%
F_0\equiv $span$\{|n,n\rangle |n=0,1,2...\}$ of the two-mode Fock space.

\section{Photon-added TMNLCS}

The photon-added quantum states were first introduced by Agarwal and Tara%
\cite{Pacs} as photon-added coherent states. Sivakumar found that the
photon-added coherent states     are NLCSs\cite{Pacsnlcs}. As a
generalization of his work, we have given a more general result that the
photon-added NLCSs are still NLCSs\cite{wang1}.

In this section, we consider the photon-added TMNLCS which is defined as

\begin{equation}
|m,n,\alpha ,f,q\rangle =\frac{a^{\dagger m}b^{\dagger n}|\alpha ,f,q\rangle 
}{\langle \alpha ,f,q|b^na^ma^{\dagger m}b^{\dagger n}|\alpha ,f,q\rangle }.
\end{equation}
Multiplying both sides of Eq.(2) by $a^{\dagger m}b^{\dagger n}$ from the
left yields 
\begin{equation}
a^{\dagger m}b^{\dagger n}f(\hat{N}_a,\hat{N}_b)ab|\alpha ,f,q\rangle
=\alpha a^{\dagger m}b^{\dagger n}|\alpha ,f,q\rangle .
\end{equation}
By the fact 
\begin{eqnarray}
a^{\dagger m}f(\hat{N}_a,\hat{N}_b) &=&f(\hat{N}_a-m,\hat{N}_b)a^{\dagger m},
\\
b^{\dagger n}f(\hat{N}_a,\hat{N}_b) &=&f(\hat{N}_a,\hat{N}_b-n)b^{\dagger n},
\end{eqnarray}
Eq.(20) can be written as

\begin{eqnarray}
&&f(\hat{N}_a-m,\hat{N}_b-n)(\hat{N}_a-m+1)(\hat{N}_b-n+1)a^{\dagger
(m-1)}b^{\dagger (n-1)}|\alpha ,f,q\rangle  \nonumber \\
&=&\alpha a^{\dagger m}b^{\dagger n}|\alpha ,f,q\rangle
\end{eqnarray}
Multiplying the both sides of the above equation by $ab$ from the left, we
get 
\begin{eqnarray}
&&f(\hat{N}_a-m+1,\hat{N}_b-n+1)(\hat{N}_a-m+2)(\hat{N}_b-n+2)  \nonumber \\
&&\times aba^{\dagger (m-1)}b^{\dagger (n-1)}|\alpha ,f,q\rangle \\
&=&\alpha (\hat{N}_a+1)(\hat{N}_b+1)a^{\dagger (m-1)}b^{\dagger
(n-1)}|\alpha ,f,q\rangle  \nonumber
\end{eqnarray}
Now we replace $m-1(n-1)$ by $m(n)$ and multiply the both sides of Eq.(24)
from the left by the operator $[(\hat{N}_a+1)(\hat{N}_b+1)]^{-1}$, the
following equation is obtained as 
\begin{eqnarray}
&&f(\hat{N}_a-m,\hat{N}_b-n)(1-\frac m{\hat{N}_a+1})(1-\frac n{\hat{N}_b+1}%
)ab|m,n,\alpha ,f,q\rangle  \nonumber \\
&=&\alpha |m,n,\alpha ,f,q\rangle
\end{eqnarray}
From Eq.(25), we can see that that the photon-added TMNLCSs are still
TMNLCSs with the corresponding nonlinear function $f(\hat{N}_a-m,\hat{N}%
_b-n)[1-m/(\hat{N}_a+1)][1-n/(\hat{N}_b+1)]$. The above discussion can be
directly generalized to the multi-mode case, but here we restrict us to the
two-mode case.

Lu and Guo studied the nonclassical properties of the photon-added pair
coherent states\cite{Papcs}. As seen from the defination of pair coherent
state and Eq.(25), we conclude that the photon-added pair coherent states
are TMNLCSs with the nonlinear function $[1-m/(\hat{N}_a+1)][1-n/(\hat{N}%
_b+1)].$

The photon-added two-mode Perelomov's coherent states\cite{Patmsvs} are
introduced and studied. From Eq.(16) and (25) we see that the photon-added
two-mode Perelomov coherent states are TMNLCSs with the nonlinear function $%
[1-m/(\hat{N}_a+1)][1-n/(\hat{N}_b+1)]/(\hat{N}_a-m+1).$

\section{Photon-subtracted TMNLCS}

By analogy to the defination of photon-added TMNLCS, the photon-subtracted
TMNLCS is defined as

\begin{equation}
|-m,-n,\alpha ,f,q\rangle =\frac{a^mb^n|\alpha ,f,q\rangle }{\langle \alpha
,f,q|b^{^{\dagger }n}a^{^{\dagger }m}a^mb^n|\alpha ,f,q\rangle }.
\end{equation}
Multiplying both sides of Eq.(2) by $a^mb^n$ from the left yields 
\begin{equation}
a^mb^nf(\hat{N}_a,\hat{N}_b)ab|\alpha ,f,q\rangle =\alpha a^mb^n|\alpha
,f,q\rangle .
\end{equation}
Using the identity 
\begin{equation}
a^mb^nf(\hat{N}_a,\hat{N}_b)=f(\hat{N}_a+m,\hat{N}_b+m)a^mb^n,
\end{equation}
we obtain

\begin{eqnarray}
&&f(\hat{N}_a+m,\hat{N}_b+n)ab|-m,-n,\alpha ,f,q\rangle  \nonumber \\
&=&\alpha |-m,-n,\alpha ,f,q\rangle .
\end{eqnarray}

From the above equation, we see that the photon-subtracted state is a TMNLCS
with the nonlinear function $f(\hat{N}_a+m,\hat{N}_b+n).$ Then
photon-subtracted pair coherent states are still pair coherent states, and
photon-subtracted two-mode Perelomov coherent states are TMNLCSs with the
nonlinear function $1/(\hat{N}_a+1+m).$

\section{Parity Pair coherent states}

By analogy to the defination in the references\cite{Spi95,Wang3}, we define
the parity operator $\Pi $ in the subspace $F_q$ as

\begin{equation}
\Pi =(-1)^{\hat{N}_b},\text{ }\Pi ^2=1,\text{ }\Pi ^{\dagger }=\Pi .
\end{equation}
Now we solve the eigenvalue equation

\begin{equation}
\Pi ab|\zeta ,q\rangle _\Pi =\zeta |\zeta ,q\rangle _\Pi .
\end{equation}
and call the state $|\zeta ,q\rangle _\Pi $ the parity pair coherent state
following the term of the parity harmonic oscillator coherent state\cite
{Spi95}. Comparing Eq.(2) and Eq.(31), we know that the parity pair coherent
state is a TMNLCS with the nonlinear function $(-1)^{\hat{N}_b}.$

From Eq.(6) and (31), the expansion of the parity pair coherent state is
easily obtained as

\begin{equation}
|\zeta ,q\rangle _\Pi =\sum_{n=0}^\infty \sqrt{\frac{q!}{n!(n+q)!}}\zeta
^n(-1)^{-n(n-1)/2}|n+q,n\rangle .
\end{equation}
The state $|\zeta ,q\rangle _\Pi $ can be rewritten as

\begin{equation}
|\zeta ,q\rangle _\Pi =\frac 1{\sqrt{2}}\left( e^{-i\pi /4}|i\zeta ,q\rangle
+e^{i\pi /4}|-i\zeta ,q\rangle \right) ,
\end{equation}
which is a superposition of two pair coherent states with phase difference $%
\pi .$ The form of this parity coherent states  is similar to that of the
parity harmonic oscillator coherent states\cite{Spi95,Yur86}.

The parity pair coherent state can be genearated in the nonlinear Kerr
medium. The Hamiltonian describing the Kerr medium is\cite{Kita86,Kral90}

\begin{equation}
H=\omega \hat{N}_b+\gamma \hat{N}_b(\hat{N}_b-1),
\end{equation}
where $\omega $ is the frequency of mode $b,$ and the parameter $\gamma $ is
proportional to the third-order nonlinear susceptibility $\chi ^3.$ The
evolution operator in the interaction picture is

\begin{equation}
U=e^{-i\gamma t\hat{N}_b(\hat{N}_b-1)}
\end{equation}
We assume the intial state is pair coherent state and let $\gamma t=\pi ,$
then the final state is just the parity pair coherent state (Eq.(32))

Analogously we can define the parity two-mode Perelomov coherent states as
the TMNLCSs with the nonlinear function $(-1)^{\hat{N}_b}/(\hat{N}_a+1)$ and
the state can be written as

\begin{equation}
|\xi ,q\rangle _\Pi =\frac 1{\sqrt{2}}\left( e^{-i\pi /4}|i\xi ,q\rangle
+e^{i\pi /4}|-i\xi ,q\rangle \right) ,
\end{equation}
which is the superposition of two two-photon Perelomov cohererent states.
Similarly the parity two-mode Perelomov coherent states can be generated in
the Kerr medium.

\section{Conclusion}

In conclusion, we have introduced the two-mode nonlinear coherent states.
The Perelomov coherent states, parity pair coherent states and parity
Perelomov coherent states are examples of the two-mode nonlinear coherent
states. An interesting  class of TMNLCSs are the photon-added or
photon-subtracted two-mode nonlinear coherent states. The corresponding
nonlinear functions are obtained for these states. Special examples,
photon-added (subtracted) pair coherent states and photon-added (subtracted)
two-mode Perelomov coherent states are discussed and the correpoding
nonlinear functions are obtained. The parity coherent states can be
generated in the Kerr medium when the correponding coherent states are the
input states. They are the superpositions of two corresponding coherent
states. The notation of the TMNLCS can be directly generalized to the
multi-mode case. The work on the multi-mode nonlinear coherent states are in
progress.

\vspace{1cm}

{\bf Acknowledgement:} The author is grateful for many valuble discussions
with Dr. Barry C. Sanders during his visit to Insitute of Physics, Chinese
Academy of Sciences. This work is supported in part by the National Science
Foundation of China with grant number:19875008.

\end{document}